\documentclass[twocolumn,runningheads]{svjour2}
\smartqed  
\usepackage{graphicx}
\journalname{Astrophysics and Space Science}
\begin{document}

\title{Status of Identification of VHE $\gamma$-ray sources
}
\titlerunning{Identification of $\gamma$-ray sources}

\author{Stefan Funk for the H.E.S.S.\ collaboration}


\institute{Stefan Funk \at Kavli Institute for Particle Astrophysics
  and Cosmology, \\ Stanford University, 2575 Sand Hill Road, PO Box 0029, Stanford, CA-94025
  USA \\ Tel.: +1-650-926-8979\\ Fax: +1-650-926-5566\\
  \email{Stefan.Funk@slac.stanford.edu} 
  / Accepted: date}

\maketitle

\begin{abstract}

 \PACS{First \and Second \and More}
\end{abstract}
With the recent advances made by Cherenkov telescopes such as
H.E.S.S.\, the field of very high-energy (VHE) $\gamma$-ray astronomy
has recently entered a new era in which for the first time populations
of Galactic sources such as e.g. Pulsar wind nebulae (PWNe) or
Supernova remnants (SNRs) can be studied. However, while some of the
new sources can be associated by positional coincidence as well as by
consistent multi-wavelength data to a known counterpart at other
wavelengths, most of the sources remain not finally identified. In the
following, the population of Galactic H.E.S.S.\ sources will be used
to demonstrate the status of the identifications, to classify them
into categories according to this status and to point out outstanding
problems.
\section{Introduction}
\label{intro}
A systematic survey of the inner part of the Galaxy performed by the
H.E.S.S.\ Cherenkov telescope system has revealed a number of
previously unknown sources of VHE gamma-rays above
100~GeV~\cite{HESSScanA}~\cite{HESSScanB}. While in terms of a
population approach the sources can be described by common properties
like generally rather hard energy spectra (photon index $\sim$2.3) or
a rather narrow distribution in Galactic latitude (rms of $\sim
0.3^{\circ}$) the counterpart identification calls for an individual
study of these objects. 
\begin{table}[t]
\caption{Categories into which the gamma-ray sources will be
  classified in the following sections.}
\centering
\label{tab:1}       
\begin{tabular}{c | c c c }
\hline\noalign{\smallskip}
 & Matching position/ & Viable emission & Consistent \\
& morphology & mechanism & MWL picture \\[3pt]
\tableheadseprule\noalign{\smallskip}
A & yes & yes & yes \\
B & no & yes & yes \\
C & yes & yes & no \\
D & no & no & no \\
\noalign{\smallskip}\hline
\end{tabular}
\end{table}
An unambiguous counterpart identification of these (initially)
unidentified H.E.S.S. sources requires {\bf (i) spatial and ideally
also morphological coincidence}, {\bf(ii) a viable gamma-ray emission
mechanism} for the object, and {\bf (iii) a consistent
multi-wavelength behaviour} matching the suggested identification and
the particle distribution within the source. The H.E.S.S.\ sources can
be classified according to their confidence in identification with
known astrophysical objects following the three requirements given
above. Table~\ref{tab:1} summarises the categories. {\bf Category~A}
comprises sources for which the positional and/or morphological match
(in case of an extended source) with a counterpart source is excellent
and the emission processes can be modelled to provide a consistent
picture describing the multi-frequency data. For these sources the
association is beyond doubt. For {\bf Category~B} sources the emission
mechanisms can be consistently modelled, however these sources show a
less convincing positional and/or morphological match with the
potential counterpart. {\bf Category C} sources on the other hand have
a good positional counterpart, they show however a non-consistent
multi-wavelength picture, being it because of insufficient data at
other wavebands, being it because of a not fully understood emission
mechanism. For {\bf Category D} sources no counterpart candidate
exists, these are the classical {\emph {unidentified sources}}. In the
following I will describe examples for sources belonging to each of
the 4 categories. The description will focus on Galactic gamma-ray
sources, since for extragalactic objects the counterpart
identification in the VHE gamma-ray regime has (so far) turned out to
be rather unproblematic.

\section{Category A - Sources with an established counterpart}
\label{subsec::CatA}
Two classes of sources can be distinguished for which a counterpart to
the VHE gamma-ray source has been established: {\bf a) point sources
with a convincing positional match} and {\bf b) extended sources with a
convincing positional and morphological match}. For these objects with a
firm counterpart, having established the positional coincidence, the
aim for these objects is to fully understand the details of the
multi-frequency photon spectrum and to investigate the emission
mechanisms generating this photon spectrum. One important question in
the VHE gamma-ray regime is for example whether the gamma-ray emission
is generated by Inverse Compton scattering of ultra-relativistic
electrons on photon fields like the Cosmic microwave background
(CMBR) or by pion-decay produced in proton-proton interactions, that
is whether the gamma-ray emission has leptonic or hadronic
origin. These two scenarios can not be directly distinguished from the
gamma-ray data alone, but have to be separated by modelling the parent
population of particles responsible for the emission. For any source
identification it should be mentioned that the good angular resolution
of VHE Cherenkov instruments (typically of the order of 0.1$^{\circ}$
per event) as well as the very low level of the diffuse gamma-ray
background at energies above 100~GeV helps against source
confusion. Source confusion was a problem that EGRET~\cite{EGRET}
strongly had to face, especially in observations in the Galactic plane
where both the density of sources and the level of the diffuse
gamma-ray background was higher. The upcoming GLAST satellite
measuring in the regime between 100~MeV to several hundreds of GeV
will have the advantage of an improved angular resolution ($\sim
0.4^{\circ}$ at 1~GeV) over EGRET but will also be susceptible to any
systematic uncertainties in modelling the diffuse gamma-ray flux from
cosmic ray interactions with interstellar material.
\begin{figure}
\centering
  \includegraphics[width=0.45\textwidth]{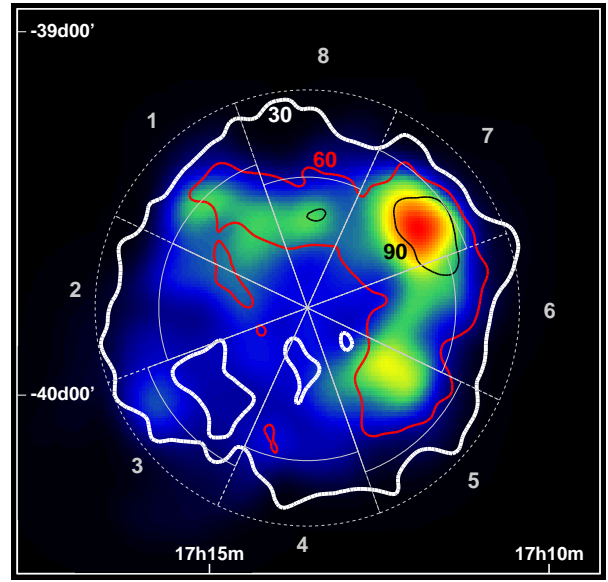}
\caption{ASCA x-ray image of RX\,J1713.7--3946, overlaid with smoothed
  and acceptance-corrected H.E.S.S.\ gamma-ray image contours linearly
  spaced at the level of 30, 60 and 90 counts. The ASCA image was
  smoothed to match the H.E.S.S.\ point-spread function for ease of
  comparison.}
\label{fig:1}       
\end{figure}
The best-established example for a {\bf point-source with a convincing
positional match} is the Crab Nebula~\cite{HESSCrab}, which is
frequently used as a calibration source in VHE $\gamma$-ray
astronomy. In order to establish a positional correlation with a
gamma-ray point-source, it has to be tested whether the nominal
position of the counterpart candidate is within the statistical and
systematic error limits of the reconstructed position of the gamma-ray
emission region (for the Crab Nebula, the statistical error on the
reconstructed position of the gamma-ray emission is 5'', the
systematic error of the order of 20''). Other objects of this class,
where a positional counterpart to a point-like gamma-ray emission has
been established are the newly discovered gamma-ray microquasars
LS\,5039~\cite{HESS5039}~\cite{HESS5039II} and
LS\,I+61\,303~\cite{MagicLSI} or the composite PWN
G0.9+0.1~\cite{HESSG0.9}. A further strengthening of the proposed
association can be established if additionally (as in the case of
LS\,5039) a correlated periodicity or variability between the
gamma-ray source and the counterpart source can be detected (LS\,5039
shows a modulation in the gamma-ray data that matches the orbital
frequency of the binary system as discussed by deNaurois et al. in
this proceedings).

The best established example for an {\bf extended source with a
convincing morphological match} is the Supernova remnant (SNR)
RX\,J1713.7--3946~\cite{HESSRXJ1713}~\cite{HESSRXJ1713II} (and also
see Lemoine-Gourmard et al. in this proceedings) showing a striking
correlation of the gamma-ray emission to X-rays as e.g. measured by
the ASCA satellite (correlation coefficient: $\sim $80\%). From the
morphological match, the association of the gamma-ray source with the
Supernova remnant is beyond doubt and can be regarded as a firm
association. Another object of this class is the Supernova remnant
RX\,J0852.0--4622 (Vela Jr.) showing also a high degree of correlation
between the gamma-ray and X-ray emission~\cite{HESSVelaJr}. Other
objects of this class are the PWNe MSH--15--5\emph{2}~\cite{HESSMSH}
and Vela~X~\cite{HESSVelaX}.

In terms of modelling the multi-frequency emission from these objects
where a firm counterpart has been established, the Crab Nebula can
again serve as an excellent example how the outstanding coverage in
wavebands from radio to VHE $\gamma$-rays can help to consistently
describe the emission mechanism in this object and to derive important
parameters like the strength of the magnetic field responsible for the
synchrotron emission. For the microquasar LS\,5039 as well as for the
extended gamma-ray emission from the Supernova remnants
RX\,J1713.7--3946 and RX\,J0852.0--4622, there is not yet a unique
solution to describe the multi-frequency data and the gamma-ray
emission can be explained both in terms of a leptonic emission
mechanism generated by Inverse Compton scattering as well as in terms
of a hadronic scenario where the gamma-ray emission is generated by
the decay of neutral pions.

\section{Category B - Sources with a less convincing positional or morphological
  match}
\label{subsec::CatB}
The best example for members of this class are the newly discovered
gamma-ray sources, which seem to belong to the so-called \emph{offset
Pulsar wind nebulae}. These objects, for which Vela~X~\cite{HESSVelaX}
is the archetypal example show an extended emission around an
energetic pulsar. The offset morphology is thought to arise from an
anisotropy in the interstellar material surrounding the pulsar, that
prevents the symmetric expansion of the PWN in one direction and
shifts the emission to the other direction (see e.g.~\cite{Blondin}
for a hydro-dynamical simulation and discussion of this effect). The
gamma-ray emission in these objects is generated by Inverse Compton
scattering of relativistic electrons accelerated in the termination
shock of the PWN. The gamma-ray sources that could possibly be
explained in this framework are typically extended and their emission
region overlaps with energetic pulsars (energetic enough to explain
the gamma-ray flux by their spindown power) and that very importantly
also show evidence for an X-ray PWN. Apart from Vela~X and
MSH--15--5{\emph2}, the gamma-ray emission of the PWN
HESS\,J1825--137~\cite{HESSJ1825}~\cite{HESSJ1825_II}, possibly powered by the
energetic pulsar PSR\,J1826--1334 can be used to illustrate the
properties of this class of objects. This object has been observed by
H.E.S.S.\ in a very deep observation of $\sim 67$ hours, due to its
proximity to the microquasar LS\,5039 (at a distance of $\sim
1^{\circ}$). It was known to be a PWN candidate also in VHE gamma-rays
since through XMM-Newton X-ray observations of
PSR\,J1826--1334~\cite{Gaensler} established an offset X-ray PWN
extending asymmetrically $\sim 5''$ to the south of the pulsar. The
asymmetric nature of the PWN as well as CO data that show a dense
molecular cloud to the north of PSR\,J1826--1334~\cite{Lemiere}
support the picture described above in which dense material to the
north shifts the PWN emission to the south. 
\begin{figure}
\centering
  \includegraphics[width=0.45\textwidth]{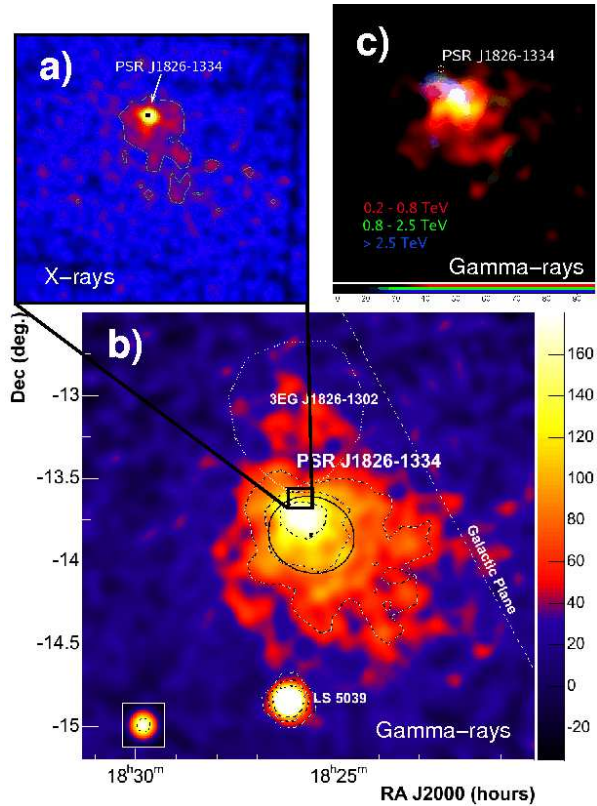}
\caption{{\bf a)} {\emph XMM-Newton} X-ray image in the energy range
  between 2 and 12 keV of the small region (7' $\times$ 7')
  surrounding PSR\,J1826--1334. It can be seen, that the X-ray
  emission shows an asymmetric diffuse emission extending to the south
  of the pulsar. {\bf b)} H.E.S.S.\ gamma-ray excess image of the
  region surrounding HESS\,J1825--137 and the energetic pulsar
  PSR\,J1826--1334 (white triangle). The image has been smoothed with
  a Gaussian of radius 2.5' and has been corrected for the changing
  acceptance across the field. The inset in the bottom left corner
  shows the PSF of the data set (smoothed in the same way as the
  excess map). The dashed black and white contours are linearly spaced
  and denote the $5 \sigma$, $10 \sigma$ and $15 \sigma$ significance
  levels. The best fit position of HESS\,J1825--137 is marked with a
  black square, the best extension and position angle by a black
  ellipse. Also shown (dotted white) is the 95\% positional confidence
  contour of the unidentified EGRET source 3EG\,J1826--1302. The
  bright point-source to the south of HESS\,J1825--137 is the
  microquasar LS\,5039.  {\bf c)} Three-colour image, showing the
  gamma-ray emission as shown in panel b), with different colours,
  denoting the gamma-ray emission in different energy bands,
  symbolising the changing morphology even in the gamma-ray band
  alone. More details can be found in the text and
  in~\cite{HESSJ1825_II}.}
\label{fig:2}       
\end{figure}

The H.E.S.S.\ detected gamma-ray emission similarly shows an
asymmetric emission extending to the south of the pulsar, however on a
completely different scale than the X-ray emission (the X-ray emission
extends 5'', whereas the gamma-ray emission extends $\sim 1^{\circ}$
to the south). This at first glance prevents a direct identification
as a counterpart, since the morphology can not be matched between
X-rays and gamma-rays. However, modelling the emission mechanism and
taking into account the different loss timescales of the gamma-ray and
X-ray emitting electrons the different scale of the emission regions
can be plausibly explained in the following way: for a typical
magnetic field of 10 $\mu \mathrm{G}$ (as also suggested from the
X-ray data), 1~keV X-rays are generated by $\sim$ 50~TeV electrons,
whereas 100~GeV gamma-rays are generated by $\sim$ 1~TeV
electrons. The gamma-rays are therefore generated by lower energy
electrons than the X-rays and the different scales of the emission
regions could be due to faster loss times of the higher energetic
synchrotron emitting electrons. This picture is further supported by
the fact, that also in the gamma-ray regime on its own, a decrease in
size with increasing energy can be established as shown
in~\cite{HESSJ1825_II} and illustrated in Figure~\ref{fig:2}.

If this picture is correct and the X-ray and gamma-ray sources are
associated, then the different morphologies in these two wavebands
(i.e. the different angular scales) can be explained in a consistent
picture. Therefore these two sources can possibly be associated, even
though there is no direct morphological match. It should be noted,
that in order to establish the association as a firm counterpart, more
multi-frequency data are needed that confirm this picture. Also it
should be mentioned, that this source generates a new template for a
large number of other extended gamma-ray sources found close to
energetic pulsars. However, in the case of HESS\,J1825--137 there are
two striking arguments that further substantiate the association: a)
the asymmetric X-ray PWN found by XMM and b) the changing morphology
found by H.E.S.S.\ in the gamma-ray data. If these two properties had
not been found, the association should not be considered more than a
chance positional coincidence. Nevertheless as mentioned, there is a
whole new class of objects that are now unidentified and that are
close to energetic pulsars, that could potential belong to this class
of objects.

\section{Category C - Sources for which the multi-frequency data does not (yet)
  provide a consistent picture}
\label{subsec::CatC}

The sources belonging to this class of objects have a good positional
counterpart, but the multi-frequency data does not (yet) provide a
consistent picture of the emission mechanism. One object of this class
is HESS\,J1813--178. This slightly extended gamma-ray source was
originally discovered in the H.E.S.S.\ Galactic plane survey and
originally described as unidentified. Shortly afterwards several
papers were published, describing the positional coincidence with an
ASCA X-ray source (2-8~keV)~\cite{Brogan}, an Integral hard X-ray
source (20-100~keV)~\cite{Ubertini}, both having the same angular
resolution and therefore like H.E.S.S.\ unable to resolve the
object. 
\begin{figure}
\centering
\includegraphics[width=0.4\textwidth]{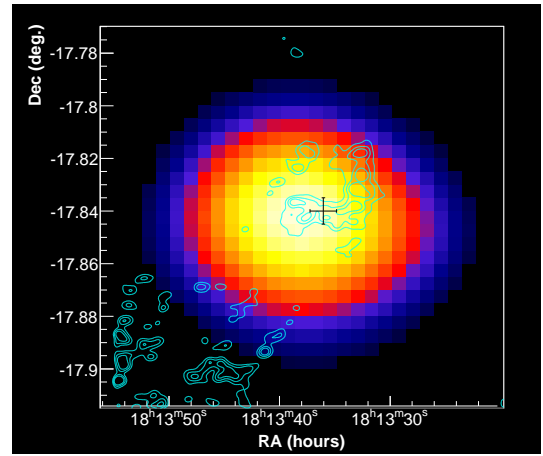}
\caption{Gamma-ray image of HESS\,J1813--178 overlaid with VLA 20 cm
  radio data in which the shell-type structure of the positional
  counterpart to the gamma-ray source can be seen. The best fit
  position of the gamma-ray excess is marked with a magenta cross.}
\label{fig:3}       
\end{figure}

Finally VLA archival radio data (90~cm) were reported~\cite{Brogan}
showing a 2.5' diameter shell-like object coincident with the X-ray
sources and with HESS\,J1813--178 (see Figure~\ref{fig:3}). This
observation led to the conclusion that the positionally coincident
object was a Supernova remnant and that the gamma-ray emission was
either caused by the shell or by a centrally located PWN. However, the
multi-frequency data does not distinguish between the two scenarios,
due to the lack of spatial resolution of the X-ray and gamma-ray
instruments. A preliminary analysis of a recent 30~ks XMM-Newton
observation of the region point to a PWN origin of the emission, which
would in turn allow to model the radio to gamma-ray data in the
picture of a composite SNR and therefore finally possibly identifying
HESS\,J1813--178 as a gamma-ray PWN. The example of HESS\,J1813--178
shows that high-quality multi-frequency are a prerequisite in the
identification of a gamma-ray source and an ongoing programme is
connected to the study of the unidentified gamma-ray sources with
high-resolution X-ray detectors like Chandra, XMM-Newton and
Suzaku. Other objects for which the multi-frequency data at this moment
do not allow firm conclusions about potential counterparts are e.g.\
HESS\,J1640--465 and HESS\,J1834--087.

\section{Category D - Sources with no counterpart} 
\label{subsec::CatD}

For this class of sources there is as of yet no counterpart at other
wavelength determined. The common belief is, that this is due to
insufficient (in terms of sensitivity) multi-wavelength data. In
principle all new gamma-ray detections are first classified in this
category, before they can be moved to another class with the help of
MWL data. The most prominent examples of this class are the HEGRA
source in the Cygnus region (TeV\,J2032+4131)~\cite{TeV2032} for which
even in deep 50~ks Chandra observation no obvious X-ray counterpart
was found. A similar case is the unidentified H.E.S.S.\ source
HESS\,J1303--631~\cite{HESS1303}, serendipitously discovered in
observations of the binary pulsar PSR\,B1259--63. Also here a Chandra
observation did not reveal any obvious positional
counterpart~\cite{Chandra1303}. In the search for X-ray counterparts
XMM-Newton seems to be best suited for the Galactic gamma-ray sources
found by H.E.S.S.\ because the high sensitivity towards extended
structures seems to be advantageous for the gamma-ray sources that
typically have extensions of the order of $0.1^{\circ}$.

A large fraction of the new H.E.S.S.\ sources can be categorised into
this class. Examples are: HESS\,J1702--420, HESS\,J1708--410 or
HESS\,J1745--303. As previously mentioned, there is an ongoing effort
to investigate these sources with various instruments from radio to
gamma-rays.

\section{Summary and Conclusion}

Table~\ref{tab:2} summarises the H.E.S.S.\ gamma-ray sources into the
proposed categories as given in table~\ref{tab:1}. As can be seen from
this table, for less than half a firm counterpart can be identified,
putting these sources into class~A. The majority of the sources has to
be classified as unidentified, since no good counterpart at other
wavebands exists at all. It is evident, that the gathering of
multi-waveband data will help in identifying these objects. It should
however also be noted, that a positional coincidence alone does not
help. A consistent modelling of the emission mechanisms at work in
generating the gamma-rays must be invoked, before a firm association
can be established. For the upcoming GLAST satellite, the situation
will be further complicated by source confusion due to the larger
point-spread function and also by the stronger diffuse gamma-ray
background from decays of neutral pions in the Galactic plane.
\begin{table}[t]
\caption{Summary of the classes for the H.E.S.S.\ Galactic gamma-ray
sources.}  \centering
\label{tab:2}       
\begin{tabular}{l | c | c }
\hline\noalign{\smallskip}
Source & Class & Comment \\[3pt]
\tableheadseprule\noalign{\smallskip}
J1713--397 & A & SNR RX\,J1713.7--3946\\
J0852--463 & A & SNR RX\,J0852.0--4622\\
J0835--455 & A & PWN Vela\,X \\
J1302--638 & A & PWN PSR\,B1259--63 \\
J1420--607 & A & PWN PSR\,J1420--6048 \\
J1514--591 & A & PWN MSH\,15--5\emph{2} \\
J1747--281 & A & PWN G0.9+0.1 \\
J1804--216 & B & PWN or SNR? \\
J1825--137 & B & PWN different scale \\
J1640--465 & C & SNR? \\
J1813--178 & C & Composite SNR? \\
J1834--087 & C & SNR? \\
J1303--631 & D & \\
J1614--518 & D & \\
J1632--478 & D & \\
J1634--472 & D & \\
J1702--410 & D & \\
J1708--420 & D & \\
J1745--290 & D & Galactic Centre source\\
J1745--303 & D & \\
J1837--069 & D & \\
\noalign{\smallskip}\hline
\end{tabular}
\end{table}

  The support of the Namibian authorities and of the University of
  Namibia in facilitating the construction and operation of H.E.S.S.\
  is gratefully acknowledged, as is the support by the German Ministry
  for Education and Research (BMBF), the Max Planck Society, the
  French Ministry for Research, the CNRS-IN2P3 and the Astroparticle
  Interdisciplinary Programme of the CNRS, the U.K. Particle Physics
  and Astronomy Research Council (PPARC), the IPNP of the Charles
  University, the South African Department of Science and Technology
  and National Research Foundation, and by the University of
  Namibia. We appreciate the excellent work of the technical support
  staff in Berlin, Durham, Hamburg, Heidelberg, Palaiseau, Paris,
  Saclay, and in Namibia in the construction and operation of the
  equipment.


 \end{document}